\documentstyle[mprocl]{article}

\def\be{\begin{equation}}
\def\ee{\end{equation}}
\def\bea{\begin{eqnarray}}
\def\eea{\end{eqnarray}}
\begin{document}

\title{BARYONIC DARK MATTER}

\author{FRANCESCO DE PAOLIS\footnote{Also at the Bartol Research
Institute, Univ. of Delaware, Newark,Delaware, 19716-4793, USA.} 
and PHILIPPE JETZER}
\address{Paul Scherrer Institute, Laboratory for Astrophysics, CH-5232 
Villigen PSI, and
Inst. of Theor. Phys., Univ. of Zurich, Winterthurerstr.
190, CH-8057 Zurich, Switzerland}

\author{GABRIELE INGROSSO}
\address{Dipartimento di Fisica and INFN, Universit\'a di Lecce,
CP 193, I-73100 Lecce, Italy}

\author{MARCO RONCADELLI}
\address{INFN, Sezione di Pavia, Via Bassi 6, I-27100 Pavia, Italy}

\maketitle
\abstracts{
Reasons supporting the idea that most of the dark matter in 
galaxies and clusters of galaxies is baryonic are discussed. 
Moreover, it is argued that most of the dark matter in galactic halos 
should be in the form of MACHOs and cold molecular clouds.}

One of the most important problems in modern astrophysics concerns the 
nature of the dark matter that 
pervades the Universe. 
%Probably, more 
%than $90\%$ of the matter in our Universe is dark.
%Evidence for the existence of dark matter comes from the observation that 
%the dynamics of many astronomical systems, such 
%as galaxies and clusters of 
%galaxies, cannot be explained by the visible matter alone. 
In the following, we discuss several reasons that lead us to 
believe that 
most of the dark matter in galaxies and clusters of galaxies 
should be baryonic. Obviously, galaxy formation remains an open problem in 
this view, and the only explanation to date requires non-baryonic dark 
matter. Still, the point we want to make is that many properties of 
galaxies and clusters of galaxies are naturally accounted for by baryonic 
dark matter alone.

{\bf MACHOs and molecular clouds}.
From the standard Big Bang nucleosynthesis model 
one infers that $0.01\le\Omega_B \le0.1$.
Since for the amount of luminous baryons 
one finds $\Omega_{\rm lum}\ll \Omega_B$, it follows that an important 
fraction of baryons are dark and they may well make  up the entire dark 
matter in galactic halos. 
Brown dwarfs and cold molecular clouds are probably the best 
candidates for dark matter in galaxies.  
Recent observations of microlensing events 
towards the Large Magellanic Clouds (LMC) 
suggest that MACHOs provide a substantial amount of the halo dark matter.
Assuming a standard spherical halo model it has been found 
that the 8 microlensing events found so far \cite{alcock}
imply a halo MACHO fraction as high as 50\% and an average mass 
of $0.27 ~M_{\odot}$ \cite{jetzer}. 
However, we note that the statistics of 
these events is at present too low to infer any definite
conclusion since both 
the halo fraction in the form of MACHOs and their average mass strongly 
depend on the assumed model for the visible and dark matter components of 
the galaxy \cite{dij}.

The problem arises of how MACHOs formed and in what form the remaining 
fraction of the galactic dark matter is. A scenario in which 
dark clusters of MACHOs and cold molecular clouds naturally form in 
the halo at large galactocentric distances has been proposed 
as well as several methods to test this model \cite{d1,d2,d3}. 
Basically, here the dynamics of the formation of dark 
clusters is similar to that of stellar globular clusters, the only 
difference being the larger galactocentric distance of dark clusters and 
consequently the lower incoming UV flux (from a central source).
This fact implies that molecular hydrogen in dark clusters is not 
dissociated so that the Jeans mass can naturally reach values as low as
$\sim 10^{-2}-10^{-1}~M_{\odot}$, leading to the formation of MACHOs.
We note that also molecular clouds should form in dark 
clusters, since the process leading to MACHO formation does not have a 
$100\%$ efficiency  and the gas cannot be expelled due to 
the absence of strong stellar winds.

Very recently, a faint optical and near-infared
emission from the halo around the galaxy NGC5907 
has been detected \cite{rudy}, which is
distributed in a manner that follows the expected distribution of the
gravitational mass and provides the first direct indication that very
faint stars with mass $\sim 0.1~M_{\odot}$ might be the repository of
most of the dark matter in the halo of galaxies.

{\bf Dark matter  at the centre of galaxies}.
Let us first assume that neutrinos make up the dark matter in
galactic halos. The 
requirement that the maximum phase-space density does not violate
the Pauli exclusion principle leads to the following lower limit for
the neutrino mass \cite{tg}:
\begin{equation}
m_{\nu}\ge 120 ~{\rm eV} \left(\frac{100~{\rm km~s}^{-1}}{\sigma}\right)^{1/4}
\left(\frac{1~{\rm kpc}}{r_c}\right)^{1/2}g_{\nu}^{-1/4}~,
\end{equation}
where $g_{\nu}$  is the number of neutrino helicity states.
For spiral galaxies ( $\sigma\sim 150$ km s$^{-1}$,  $r_c\sim 10$
kpc) one gets $m_{\nu}\ge 25$ eV; for bright elliptical
galaxies one gets $m_{\nu}\ge 5-7$ eV. 
However, when considering
dwarf galaxies ($\sigma\sim 20$ km s$^{-1}$ and $r_c\sim 1$ kpc) one
gets $m_{\nu}\ge 200$ eV, which is clearly in contraddiction with the
cosmological bound.

As next we consider cold dark matter as a candidate for the
dark matter in galactic halos. 
In this respect, several computer simulations of the large scale 
structure  with a sufficiently high resolution 
to resolve the internal structure of the galactic halos, seem to 
indicate that the density profiles of the cold dark matter
should have central cusps. 
These cusps are incompatible with the isothermal density profiles
$\rho(r)={\rho(0)}/[1+(r/r_c)^2]$.
While this profile becomes 
approximately constant at $r\ll a$ and has a finite central density 
$\rho(0)$, numerical simulations \cite{burkert} 
indicate a density distribution that diverges like $r^{-1}$. 
The existence of a central density cusp in 
normal galaxies is difficult to demonstrate since the internal regions 
are gravitationally dominated by the visible component. 
On the contrary, dwarf spiral galaxies provide excellent 
probes for the internal structure of dark halos since 
these galaxies are completely dominated by dark 
matter on scales larger than a kiloparsec \cite{cf}.
One can, therefore, use these galaxies to 
investigate the inner structure of 
dark halos with very little ambiguity about the contribution from the 
luminous matter and the resulting uncertainties in the disk 
mass/luminosity ratio (M/L).
Only about a dozen rotation curves of dwarf galaxies have been measured, 
but a trend clearly emerges: 
the rotational velocities rise over most of the 
observed region, which spans several times the optical scale lengths and
nevertheless lies within the core radius of the mass distribution.
Rotation curves of dwarf galaxies  do not admit singular density 
profiles at the galactic centre and their profiles are in good agreement 
with the isothermal density law.
Given the above considerations, we conclude that the dark matter in
dwarf galaxies has to be mainly baryonic and therefore, it is very
likely that also in ellipticals and spirals it should
be baryonic as well.

{\bf Galactic evolution along the Hubble sequence}.
It has been pointed out \cite{pcm} that spiral galaxies 
evolve along the Hubble sequence from $S_d$ to $S_a$ in  billions of 
years. During this evolution the dimensions of both galactic nuclei 
and disks increase while the M/L ratio should decrease. This fact
suggests that dark matter gradually transforms into visible matter, 
that is in stars. 
Of course, this is possible only if the dark matter is 
baryonic and, in particular, if it is in gaseous form.

{\bf Rotation curve shapes}.
Initial studies have indicated that
rotation curves of spiral galaxies are generally flat. 
This means that the galactic halo 
must produce practically the entire 
rotational velocity far out the optical 
radius, while in the internal regions the optical disk maximally 
contributes to the rotation curve. It seems that disk and halo 
combine together to produce a flat rotation curve. This synthony between 
disk and halo has been called the {\it disk-halo conspiracy}.
However, this {\it conspiracy} is not always true. 
In some dwarf galaxies the dark 
halo mass is considerably higher than the luminous disk mass inside the 
optical radius. 
In the internal regions of bright spirals the disk is the dominant 
component, while the halo contributes significantly to the rotation curve 
only at large galactocentric distances \cite{cv}. 
The dependence of the rotation curves on the luminous content of the spiral 
galaxies, we are talking about, can be explained if the dark matter in 
spirals is baryonic and in 
particular if halos formed before galactic disks, 
as it naturally happens in our model \cite{d2}.

{\bf Dark matter in clusters of galaxies}.
It is well known that the 
ratio M/L increases from the luminous part of galaxies to clusters 
and superclusters of galaxies. This fact has 
generally induced astrophysicists to conclude that clusters and 
superclusters of galaxies have much more matter per unit luminous matter 
than individual galaxies, so that the critical density of the Universe 
can be attained.
Recently, it has been shown \cite{bld}  that most of the 
dark matter in clusters and superclusters of galaxies should be clumped in the 
halos around galaxies. Indeed, the ratio M/L in clusters does not 
significantly increase at scales larger than 100--200 kpc, typical of 
galactic halos. The total mass of the clusters can then be accounted for 
by the mass of the galaxies plus the hot gas mass 
($\sim 20\%$ of the total cluster mass). 
This, in addition to the fact that the voids seem to give a minor contribution
to $\Omega$, suggests that  $\Omega$ can be as low as 
$\sim 0.2$.

%{\bf Lyman-$\alpha$ absorption systems}.
%It is well known that Quasar Ly-$\alpha$ absorption lines provide a
%detailed information on the evolution of the gaseous component of 
%galaxies. 
%It looks natural to identify Ly-$\alpha$ forest 
%clouds with the molecular clouds clumped into dark clusters located 
%in the outer halo. We expect that the clouds contain
%in their external layers an increasing fraction of $HI$ gas and that the outer
%regions are even ionized, due to the incoming UV radiation.
%A $HI$ column density of $\sim 10^{14}$ cm$^{-2}$ -- corresponding 
%to a layer of about $10^{-6}$ pc -- is sufficient to shield the 
%incoming radiation. 
%Thus, since we can estimate that typically $\sim 100$
%clouds are intercepted
%along the line of sight to a distant Quasar, we get a value
%of $\sim 10^{16}$ cm$^{-2}$, which is
%consistent with the column density observed for the Ly-$\alpha$ forest clouds.
%As a final comment, we mention that recently R\"ottgering et al. 
%\cite{rmv}, by considering the filling factors and the 
%physical parameters 
%derived from their Ly-$\alpha$ forest observations, pointed out that 
%galactic halos may contain $\sim 10^9~M_{\odot}$ of 
%neutral hydrogen gas and are 
%typically composed of $\sim 10^{12}$ clouds, each of size $\sim$ 
%40 light-days. 

\end{document}